\documentclass[superscriptaddress, prl, twocolumn,showpacs]{revtex4}

\usepackage{amsmath}
\usepackage{amssymb}
\usepackage{epsfig}
\usepackage{graphicx}
\usepackage{wasysym}
\usepackage{bbm}

\newcommand \be{\begin{equation}}
\newcommand \ee{\end{equation}}
\newcommand \bes{\begin{equation*}} 
\newcommand \ees{\end{equation*}}
\newcommand \bea{\begin{eqnarray}}
\newcommand \eea{\end{eqnarray}}
\newcommand \bsea{\begin{subequations}\begin{eqnarray}} 
\newcommand \esea{\end{eqnarray}\end{subequations}}
\newcommand \beas{\begin{eqnarray*}} 
\newcommand \eeas{\end{eqnarray*}}
\newcommand \bfg{\begin{figure}}
\newcommand \efg{\end{figure}}
\newcommand \bfgs{\begin{figure*}} 
\newcommand \efgs{\end{figure*}}
\newcommand \bwt{\begin{widetext}}
\newcommand \ewt{\end{widetext}}
\newcommand \im{\mathbbm{i}} 
\newcommand \nd{{\vphantom{\dagger}}} 
\newcommand \bra{\langle}
\newcommand \ket{\rangle}

\newcommand \vJ{\mathbf{J}}
\newcommand \vB{\mathbf{B}}
\newcommand \vL{\mathbf{L}}
\newcommand \vM{\mathbf{M}}
\newcommand \vS{\mathbf{S}}
\newcommand \vq{\mathbf{q}}
\newcommand \vQ{\mathbf{Q}}

\newcommand \bt{\bar{t}}

\begin{document}

\title{
Twisted Hubbard Model for Sr$_2$IrO$_4$: 
Magnetism and Possible High Temperature Superconductivity.
}

\author{Fa Wang}
\affiliation{Department of Physics, Massachusetts Institute of Technology, 
Cambridge, Massachusetts 02139}
\author{T. Senthil}
\affiliation{Department of Physics, Massachusetts Institute of Technology, 
Cambridge, Massachusetts 02139}

\date{\today}

\begin{abstract}
Sr$_2$IrO$_4$ has been suggested as a Mott insulator 
from a single $J_{\rm eff}=1/2$ band, similar to the cuprates. 
However this picture is complicated by 
the measured large magnetic anisotropy and ferromagnetism.
Based on a careful mapping to the $J_{\rm eff}=1/2$(pseudospin-1/2) space, 
we propose that 
the low energy electronic structure of Sr$_2$IrO$_4$ can indeed be described 
by a SU(2) invariant pseudospin-1/2 Hubbard model 
very similar to that of the cuprates, 
but with a ``twisted'' coupling to external magnetic field 
(a $g$-tensor with a staggered antisymmetric component).
This perspective naturally explains the magnetic properties of Sr$_2$IrO$_4$.
We also derive several simple facts based on this mapping
and the known results about the Hubbard model and the cuprates, 
which may be tested in future experiments on Sr$_2$IrO$_4$.
In particular we propose that (electron-)doping Sr$_2$IrO$_4$ 
can potentially realize high-temperature superconductivity. 
\end{abstract}


\maketitle


Various Ir oxides have recently become the platform to study 
the interplay between strong spin-orbit(SO) interaction and 
strong correlation effects. 
There has been an experimental observation of a three-dimensional 
spin liquid in a hyper-kagome structure of Na$_4$Ir$_3$O$_8$\cite{hyperkagome}. 
Theoretical proposals such as  
the realization of correlated topological insulators\cite{correlatedTI},
the Kitaev model\cite{Khaliullin}, 
and a Dirac semimetal with surface ``Fermi arcs''\cite{Ashvin} 
in iridates have been made as well. 
Here we propose that doped Sr$_2$IrO$_4$ may realize 
high-temperature superconductivity similar to the cuprates. 


The crystal structure of Sr$_2$IrO$_4$ 
consists of two-dimensional(2D) IrO$_2$ layers, 
similar to 
the parent compound La$_2$CuO$_4$ of the cuprates. 
The main difference is that the oxygen octahedra surrounding Ir 
rotate along $c$-axis by about $11^\circ$ in a staggered pattern, 
enlarging the unit cell by $\sqrt{2}\times\sqrt{2}\times 2$ 
\cite{structure}. 
The electronic structure of Sr$_2$IrO$_4$ is quasi-2D, 
but is expected to have several differences from the cuprates. 
Ir$^{4+}$ has the electronic structure $5d^{5}$, 
so the $t_{2g}$ levels should to be active,  
while Cu$^{2+}$ with $3d^{9}$ configuration has only 
the top $e_{g}$ level active. 
Ir as a $5d$ transition metal is expected to have weaker 
correlation effects than $3d$ elements(e.g. Cu). 
At this point one may expect that Sr$_2$IrO$_4$ is 
a (multi-band) weakly correlated metal. 
But strong spin-orbit coupling of Ir dramatically changes the story. 
The $t_{2g}$ levels are split by SO interactions into 
a higher energy Kramers doublet 
(the pseudospin-1/2 or $J_{\rm eff}=1/2$ states)
and two pairs of lower energy ones\cite{Bleaney}. 
These $J_{\rm eff}=1/2$ states are equal weight superpositions of 
all three $t_{2g}$ orbitals,
and this has been confirmed experimentally
by resonant x-ray scattering\cite{xray}
and theoretically by LDA+SO+U calculation\cite{LDA}. 
With $d^5$ configuration of Ir the $J_{\rm eff}=1/2$ states 
are half-filled. 
They have much smaller band width than expected for  
the $t_{2g}$ levels without SO interaction and
therefore have effectively enhanced correlation effect. 
In the end Sr$_2$IrO$_4$ is a Mott insulator
and exhibits magnetic order below 240K\cite{resistivity,optical,ARPES-optical}. 

It is then tempting to make the analogy between Sr$_2$IrO$_4$ and the cuprates
and speculate that doped Sr$_2$IrO$_4$ can also realize 
the interesting physics in doped cuprates, 
{e.g.} superconductivity, pseudogap, stripe formation, {etc.}. 
But strong SO interaction, different active orbitals and 
the rotation of oxygen octahedra seem to significantly complicate the problem. 
For example, Sr$_2$IrO$_4$ has very anisotropic susceptibility
and shows ferromagnetism(FM) 
with large ferromagnetic moment $\sim 0.14\mu_{B}$ per Ir\cite{ferromag}, 
which was attributed to Dzyaloshinskii-Moriya(DM) interaction generated by 
the rotation of oxygen octahedra. However it has been pointed out 
by Jackeli and Khaliullin\cite{Khaliullin} that 
the DM interaction can be removed by staggered rotation of pseudospin space 
on Ir sites.
We will extend this consideration to the electronic model 
in a slightly different perspective and 
discuss more details and consequences of this mapping. 
Finally we will argue that the analogy between Sr$_2$IrO$_4$ and the cuprates
can be established with careful interpretation, and 
that interesting physics of doped cuprates may also be realized 
in doped Sr$_2$IrO$_4$.

{\bf The mapping to one band Hubbard model.}
\begin{figure}
\includegraphics[scale=0.2]{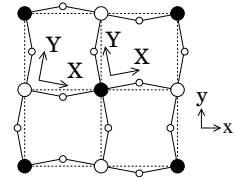}
\caption{
Schematic picture of one IrO$_2$ layer. Large filled/open circles 
indicate the Ir atoms on two sublattices. Small open circles are oxygens.
Small $x,y$ are the global axis, while captial $X,Y$ indicate local 
cubic axis(sublattices dependent).
}
\label{fig:1}
\end{figure}
To begin with we will treat Sr$_2$IrO$_4$ as quasi-2D 
and consider only one IrO$_2$ layer, which is schematically illustrated in
 Fig.~\ref{fig:1}.
Label the rotation angle of oxygen octahedron around Ir site $j$ 
by $\theta_{j}=\epsilon_{j}\theta$, 
with $\epsilon_{j}=\pm 1$ for the two sublattices and
 $\theta\approx 11^\circ$ from experiments\cite{structure}. 
The crystal-field splitting of $t_{2g}$ and $e_{g}$ levels 
and projection to $J_{\rm eff}=1/2$ states should be implemented in
the rotated local cubic axis. 
Label the global axis by $x,y,z$ and local cubic axis (on site $j$) 
by $X,Y,Z$ (see Fig.~\ref{fig:1}).
The unit vectors of these two coordinates systems are related by 
\be
\hat{X}=\hat{x}\,\cos \theta_{j}+\hat{y}\,\sin \theta_{j},\ \ 
\hat{Y}=-\hat{x}\,\sin \theta_{j}+\hat{y}\,\cos \theta_{j},\ \ 
\hat{Z}=\hat{z}.
\label{equ:XYZxyz}
\ee

The $J_{\rm eff}=1/2$ states are
(see {e.g.} Ref.~\cite{Bleaney}, 
the phase convention here is slightly different,
$\im=\sqrt{-1}$)
\be
\begin{split}
&
|J_{\rm eff}^z=+1/2\ket
=\frac{1}{\sqrt{3}}\left ( 
+\im |XY,\uparrow\ket
-|XZ,\downarrow\ket
+\im |YZ,\downarrow\ket
\right ),
\\ &
|J_{\rm eff}^z=-1/2\ket
=\frac{1}{\sqrt{3}}\left ( 
-\im |XY,\downarrow\ket
+|XZ,\uparrow\ket
+\im |YZ,\uparrow\ket
\right ).
\end{split}
\label{equ:Jstates}
\ee
$XZ,YZ,XY$ are the $t_{2g}$ orbitals defined in the {\em local} cubic axis. 
$\uparrow,\downarrow$ indicate spin states
(defined also in the {\em local} cubic axis).
Note that although the elongation of oxygen octahedra along
 $c$-axis is expected to 
change the relative weights of the three orbitals\cite{Bleaney,Khaliullin}, 
this has not been observed in resonant x-ray scattering experiment\cite{xray}
or LDA+SO+U calculation\cite{LDA}. 

As the first approximation, the effective electronic Hamiltonian should
be the projection of full Hamiltonian on the 
subspace of $J_{\rm eff}=1/2$ states. 
Considering first the Hamiltonian on the $t_{2g}$ subspace, 
we expect the following, 
1) the $t_{2g}$ orbitals should be defined in 
 the {\em local} cubic axis basis, 
 because the crystal-field on Ir $5d$ orbitals from neighboring oxygens is 
 diagonal only in the {\em local} cubic axis;
2) assuming that hoppings between Ir sites are mediated by 
 the oxygen $2p$ orbitals, simple symmetry consideration shows 
 that effective hoppings between nearest-neighbor Ir 
 are orbital diagonal(one $t_{2g}$ orbital does not hop to another orbital) 
 only in the {\em local} cubic axis basis; 
3) if the spin spaces are defined in the {\em global} axis basis, 
 the effective hoppings of Ir $t_{2g}$ orbitals will be real. 
Two tight-binding models on the $t_{2g}$ subspace 
have been obtained by fitting LDA+SO+U dispersions  
in Ref.~\cite{LDA} and Ref.~\cite{VMC}, 
and both have this property of real orbital diagonal hoppings,
but no clear interpretation was given. 
The discussion above shows that the orbitals in these models 
should be interpreted as the $t_{2g}$ orbitals in the {\em local} cubic axis, 
while the spins in these models are defined in the {\em global} axis. 

The spin space on every site should be first rotated to {\em local} axis
before the projection to the $J_{\rm eff}=1/2$ states, 
because the spins used in (\ref{equ:Jstates}) are defined in local axis. 
Namely we need to interpret the electron operators
 $c_{j,a,\nu}^\dagger$ used in these models,
 on site $j$ for orbital $a=XZ,YZ,XY$ with spin $\nu$, 
 as creation operators for the states
 $e^{\im \epsilon_{\nu}\theta_{j}/2} |j,a,\nu\ket$,
 where $\epsilon_{\nu}=\pm 1$ for spin index $\nu=\uparrow,\downarrow$
 respectively. 

Define $d_{\uparrow}$ and $d_{\downarrow}$ as the annihilation operators 
for the $|J_{\rm eff}^z=\pm 1/2\ket$ states (\ref{equ:Jstates}) respectively. 
The projection on the $J_{\rm eff}=1/2$ subspace 
is implemented by the following substitution, 
$
c_{j,XY,\nu}^\dagger \to
 -\epsilon_{\nu}\im \sqrt{1/3}\, e^{\im\epsilon_{\nu}\theta_{j}/2} d_{j,\nu}^\dagger$,
$
c_{j,XZ,\nu}^\dagger \to
 \epsilon_{\nu} \sqrt{1/3}\, e^{\im\epsilon_{\nu}\theta_{j}/2} d_{j,-\nu}^\dagger$,
and $
c_{j,YZ,\nu}^\dagger \to
 -\im \sqrt{1/3}\, e^{\im\epsilon_{\nu}\theta_{j}/2} d_{j,-\nu}^\dagger$.
The onsite interactions between $t_{2g}$ orbitals will be projected 
into an onsite $U$ term of the Hubbard model
for the $J_{\rm eff}=1/2$ states
due to time-reversal symmetry and charge conservation. 

We take as a concrete example the tight-binding model of Ref.~\cite{VMC}. 
It involves nearest-neighbor(NN) $XY$ hopping $t_1=0.36$eV,
 NN $XZ$($YZ$) hopping along $x$($y$) direction $t_4=0.37$eV, 
 NN $XZ$($YZ$) hopping along $y$($x$) direction $t_5=0.06$eV, 
 next-nearest-neighbor $XY$ hopping $t_2=0.18$eV,
 and third-neighbor $XY$ hopping $t_3=0.09$eV. 
The resulting one band Hubbard model after projection is
\be
\begin{split}
H=
\ &
-\sum_{<jk>,\alpha}(t+\im\epsilon_{\alpha}\epsilon_{j}\bt)
 \, d_{j,\alpha}^\dagger d_{k,\alpha}^\nd
-\sum_{<<jk>>,\alpha}t'\, d_{j,\alpha}^\dagger d_{k,\alpha}^\nd
\\ & 
-\sum_{<<<jk>>>,\alpha}t''\, d_{j,\alpha}^\dagger d_{k,\alpha}^\nd
+U\sum_{j}d_{j,\uparrow}^\dagger d_{j,\uparrow}^\nd 
 d_{j,\downarrow}^\dagger d_{j,\downarrow}^\nd
\end{split}
\label{equ:tttUmodel}
\ee
with $\alpha=\uparrow,\downarrow$, 
and the effective hoppings are
$t =(1/3)(t_1+t_4+t_5)\cos\theta\approx 0.258{\rm eV}$,
$\bt =(1/3)(t_1-t_4-t_5)\sin\theta\approx -0.0045{\rm eV}$,
$t' =(1/3)t_2\approx 0.06{\rm eV}$,
$t'' =(1/3)t_3\approx 0.03{\rm eV}$.
$\bt$ is very small and will be ignored hereafter. 
In general $\bt$ can be absorbed into $t$ by a unitary transformation
 $d_{j,\alpha}\to e^{\im \epsilon_{\alpha} \epsilon_{j} \phi/2}
 \tilde{d}_{j,\alpha}$
with $\phi={\rm arc}\tan(\bt/t)$, but we will not elaborate on this. 
The value of $U$ has been estimated as
 $\sim 2{\rm eV}$\cite{LDA,ARPES-optical}. 
This $t-t'-t''-U$ model has been widely used as an effective model 
for the cuprates, 
although the parameters here have different values. 

With large $U$ and at half-filling the model (\ref{equ:tttUmodel}) 
is an Mott insulator described by a pseudospin-1/2 model with SU(2) symmetry. 
If second- and third-neighbor $t',t''$ are ignored  
the half-filling pseudospin model to the lowest order of $t/U$ 
is just the Heisenberg AFM model of pseudospins $\vJ$, 
$H_{AFM}= \sum_{<jk>}(4t^2/U)\vJ_{j}\cdot\vJ_{k}$.
Each pseudospin has three components ($a=1,2,3$) 
$
J_{j,a}=(1/2)\sum_{\alpha,\beta}d_{j,\alpha}^\dagger
 (\sigma^{a})_{\alpha\beta}^\nd d_{j,\beta}^\nd
$,
where $\sigma$ are Pauli matrices
and $\alpha,\beta=\uparrow,\downarrow$ label 
the $J_{\rm eff}^z=\pm 1/2$ states.

{\bf Coupling to external magnetic field.} 
Although the effective model (\ref{equ:tttUmodel}) looks 
exactly like the model of the cuprates, 
the coupling to external magnetic field in Sr$_2$IrO$_4$ is 
quite different. 

Assume the coupling of magnetic field $\vB$ on Ir $5d$ orbitals 
is described by the atomic form
(more careful treatment can be found in, {e.g.}, Ref.~\cite{Bleaney}), 
 $H_{B}=-\mu_B \vB\cdot(\vL+2\vS)$,  
where $\mu_B$ is the Bohr magneton. 
After projection to the $J_{\rm eff}=1/2$ states it becomes 
 $H_{B}=2\mu_B \vB\cdot\vJ=2\mu_B(B_X J_1+B_Y J_2+B_Z J_3)$. 
Note that $B_{X,Y,Z}$ are components of field on the {\em local} cubic axis,
 $B_X=\vB\cdot\hat{X}$ etc.. 
Use the relation (\ref{equ:XYZxyz}),
the coupling on site $j$ in terms of the field components on the global axis,
$B_x,B_y,B_z$, is

\be
\begin{split}
H_{B,j}=
2\mu_B[
& 
 B_{j,x}(J_{j,1}\cos\theta_{j}-J_{j,2}\sin\theta_{j})
\\ & 
+B_{j,y}(J_{j,2}\cos\theta_{j}+J_{j,1}\sin\theta_{j})
+B_{j,z}\,J_{j,3}.
]
\end{split}
\ee

Therefore the observable magnetic moment $\vM_{j}$ on site $j$ has 
the following components on the global axis, 
\be
\begin{pmatrix}
M_{j,x} \\
M_{j,y} \\
M_{j,z}
\end{pmatrix}
= -2\mu_B 
\begin{pmatrix} 
\cos\theta & -\epsilon_{j}\sin\theta & 0 \\
\epsilon_{j}\sin\theta & \cos\theta & 0 \\
0 & 0 & 1
\end{pmatrix}
\begin{pmatrix}
J_{j,1} \\
J_{j,2} \\
J_{j,3}
\end{pmatrix}.
\label{equ:MJ}
\ee
If $\bt$ in (\ref{equ:tttUmodel}) is not ignored 
$\theta\approx 11^\circ$ in (\ref{equ:MJ}) should be replaced by
 $\theta-{\rm arc}\tan(\bt/t)\approx 12^\circ$. 
This nontrivial relation between moments $\vM$ and pseudospins $\vJ$, 
namely a $g$-tensor with a staggered antisymmetric component, 
has several interesting consequences which we list below. 

\begin{itemize}
\item
By quantum Monte Carlo studies\cite{QMC-square} 
the square lattice Heisenberg model has a N\'eel ground state 
with staggered ``magnetization''
$|\bra \epsilon_{j}\vJ_{j}\ket|\approx 0.307$. 
However because of the relation (\ref{equ:MJ}), 
the ordered moments do not form a simple collinear N\'eel pattern. 
If the ordered moments lie in the $xy$ plane,
they will be rotated together with the oxygen octahedra 
in a staggered pattern   
therefore create a net ferromagnetic moment per site, 
$2\mu_B\cdot |\bra \epsilon_{j}\vJ_{j}\ket|\cdot\sin \theta \approx 0.12\mu_B$.
This is very close to the experimentally observed value $0.14\mu_B$ 
\cite{ferromag}.

\item
By the relation (\ref{equ:MJ}) we can relate pseudospin correlation
functions of model (\ref{equ:tttUmodel}) 
to moment correlation functions which is actually measured by 
susceptibility or magnetic neutron/x-ray scattering experiments. 
The Fourier components of moments with wavevector $\vq$
and frequency $\omega$ is related to pseudospins by,
\bes
\begin{split}
&
M_{\vq,\omega,x} = -2\mu_B[\cos \theta\, J_{\vq,\omega,1}
 -\sin \theta\, J_{\vq+\vQ,\omega,2}],
\\ &
M_{\vq,\omega,y} = -2\mu_B[\cos \theta\, J_{\vq,\omega,2}
 +\sin \theta\, J_{\vq+\vQ,\omega,1}],
\\ &
M_{\vq,\omega,z} = -2\mu_B\, J_{\vq,\omega,3}.
\end{split}
\ees
where $\vQ=(\pi,\pi)$ is the wavevector of N\'eel order.
In the paramagnetic phase the dynamical susceptibility $\chi^{ab}(\vq,\omega)$,
which is proportional to the ``moment structure factor''
 $\bra M_{\vq,\omega,a}M_{-\vq,-\omega,b}\ket$,
is related to the dynamical pseudospin susceptibility
 $\chi^{ab}_{J}(\vq,\omega)=\delta_{ab}\chi_{J}(\vq,\omega)
 \propto \bra \vJ_{\vq,\omega}\cdot\vJ_{-\vq,-\omega}\ket$
by
\bes
\begin{split}
&
\chi^{xx}(\vq,\omega)=\chi^{yy}(\vq,\omega) 
\\
=\ &
 \cos^2\theta\,\chi_{J}(\vq,\omega)
+\sin^2\theta\,\chi_{J}(\vq+\vQ,\omega),
\end{split}
\ees
$\chi^{zz}(\vq,\omega)=\chi_{J}(\vq,\omega)$, 
and other components of $\chi^{ab}$ are zero. 
In particular the measured static uniform ($\omega=0,\vq=0$) susceptibility 
in $xy$ plane is actually a mixture of 
the uniform and staggered susceptibility of the SU(2) invariant 
Hubbard/Heisenberg model. 
This explains in a different perspective 
the measured large anisotropy of uniform susceptibility 
and the ferromagnetic Curie-Weiss law \cite{ferromag}. 
In our picture the anisotropy is not mainly from easy axis interaction
suggested by Ref.~\cite{ferromag}
but from the mixing of large staggered susceptibility,  
and the FM Curie-Weiss law comes from 
the contribution of staggered susceptibility close to 
AFM N\`eel order of pseudospins.

\item
In the high temperature paramagnetic phase above  
the N\'eel ordering temperature, 
the measured moment-moment correlation will be dominated by 
the staggered pseudospin correlation of a SU(2) invariant model, 
although the measured susceptibility shows significant anisotropy. 
The moment-moment correlation length will behave like 
the 2D Heisenberg model\cite{CHN}, 
which has recently been observed by magnetic x-ray scattering
 \cite{correlationlength}.

\end{itemize}

{\bf Possible high-temperature superconductivity.}
If the one band Hubbard model (\ref{equ:tttUmodel}) 
is indeed a good approximation of the electronic struture of Sr$_2$IrO$_4$,
and if the high-temperature superconductivity in doped cuprates 
is indeed described by the one band Hubbard model,
a natural consequence is that doped Sr$_2$IrO$_4$ will realize
high-temperature superconductivity. 
In the following we list several direct consequences from this analogy. 

\begin{itemize}

\item
It is believed that the sign and magnitude of $t'$ is important for high-Tc
in the cuprates 
and likely responsible for the particle-hole asymmetry of the phase diagram.
(see {e.g.}, Ref.~\cite{RMP}). 
The relative magnitude $|t'/t|\approx 0.23$ for Sr$_2$IrO$_4$ is similar to 
the cuprates. 
However the sign of $t'$ for Sr$_2$IrO$_4$ is opposite to that of the cuprates. 
This can be remedied by a particle-hole transformation
 $d_{j,\alpha}^\nd\to \epsilon_{j}^\nd d_{j,\alpha}^\dagger$. 
Therefore we expect that the doping phase diagram of Sr$_2$IrO$_4$ 
will be the particle-hole conjugate of the cuprates, in particular  
high-Tc will be easier to achieve on the electron-doped side 
of Sr$_2$IrO$_4$, 
{e.g.} with La substitution of Sr. 
Interestingly electron-doped Sr$_2$IrO$_{4-\delta}$ has recently been 
synthesized and metallic behavior was reported for $\delta=0.04$ \cite{dope}.

\item
The interlayer hopping of the cuprates is of the form
 $t_{\perp}(k_{\parallel})=t_{\perp 0}\, v^2$
with $v=(\cos k_x-\cos k_y)/2$, due to
the $d_{x^2-y^2}$ orbital content\cite{interlayer}. 
This together with the $d_{x^2-y^2}$ nodal pairing symmetry
significantly suppress transport along $c$-axis, 
making the superconducting properties of the cuprates very anisotropic. 
However the resistivity anisotropy $\rho_{c}/\rho_{ab}$ of Sr$_2$IrO$_4$
is only $10^2-10^3$ \cite{magnetoelectric},
very small compared to $10^4-10^5$ of the cuprates\cite{cuprateanisotropy}, 
which implies a larger $t_{\perp 0}$ for Sr$_2$IrO$_4$. 
The active orbitals for Sr$_2$IrO$_4$ is very different from the cuprates and 
the factor $v^2$ should be different and not vanish on the nodal direction. 
Both facts suggest that 
Sr$_2$IrO$_4$ should have more isotropic superconducting properties
which is beneficial for practical applications. 

\item
The pairing will be a pseudospin singlet $d_{x^2-y^2}$ pairing 
and in many ways behave like the d-wave pairing of the cuprates. 
Phase sensitive and other indirect measurements used to determine the 
d-wave symmetry in the cuprates can be applied to doped Sr$_2$IrO$_4$ as well. 

\item
The energy scale of the one band Hubbard model for Sr$_2$IrO$_4$
is lower than that of the cuprates by about $50\%$. 
Therefore the $T_c$ of doped Sr$_2$IrO$_4$ will likely be lower than 
the cuprates.

\end{itemize}

{\bf Discussion and Conclusion.} 
The one band Hubbard model (\ref{equ:tttUmodel}) is of course 
the zeroth order approximation of the low energy electronic 
structure of Sr$_2$IrO$_4$. 
In real material the mixing between the $J_{\rm eff}=1/2$ states
and other states will generate anisotropy in pseudospin interactions, 
which will be important close to and below the N\'eel temperature. 
For magnetic properties above the N\'eel temperature and 
for electron-doped Sr$_2$IrO$_4$ we believe this one band Hubbard model is 
still a good description. 

The projection to one band Hubbard model was also implemented in
Ref.~\cite{LDA}. 
The resulting hoppings reported in Equ.~(7)(8) of Ref.~\cite{LDA}
suggest that the authors of Ref.~\cite{LDA} 
interpreted the orbitals in their $t_{2g}$ tight-binding model 
as the {\em global} axis basis $xz,yz,xy$. 
Here we have argued that the orbitals should be interpreted as 
the {\em local} axis basis 
which produces a projection result 
[$\bar{t}_0=-(2 t_0/3)\cos\theta$ and $\bar{t}_1=0$] 
different from Ref.~\cite{LDA}.

In summary we have performed the projection of the electronic structure
of Sr$_2$IrO$_4$ to the $J_{\rm eff}=1/2$ states and carefully 
deduced the resulting one band Hubbard model and its interpretation. 
We provide another perspective on the magnetic properties of Sr$_2$IrO$_4$
by viewing it as a SU(2) invariant Hubbard/Heisenberg pseudospin-1/2 model,
but with a twisted relation (\ref{equ:MJ}) between 
the observable moments and the pseudospin degrees of freedom, 
namely a $g$-tensor with staggered antisymmetric component. 
One direct consequence is that the measured uniform susceptibility 
in $ab$ plane is actually a mixture of uniform 
and staggered susceptibility of SU(2) invariant Hubbard/Heisenberg model. 
Despite the complication of strong SO interaction, different active orbitals 
and structure distortion, 
the effective one band Hubbard model of Sr$_2$IrO$_4$ remarkably resembles 
the cuprates.
We thus propose that doped Sr$_2$IrO$_4$ can realize high-temperature 
superconductivity, and potentially other interesting physics 
of the cuprates.
By comparing the model parameters 
we suggest that electron-doping of Sr$_2$IrO$_4$ will be the analogue of 
hole-doping of the cuprates.
This can be achieved by La substitution of Sr, or O deficiency \cite{dope}, 
and maybe by field effect on thin films\cite{fieldinducedSC}, 
 or interfacing with other oxides\cite{interface}.
We hope these simple theoretical observations will stimulate more 
experimental research on Sr$_2$IrO$_4$. 

{\bf Acknowledgment} 
The authors thank Leon Balents, Dung-Hai Lee, Patrick A. Lee, 
and Michael Norman for helpful discussions.
TS was supported by NSF through the grant DMR-1005434.

\end{document}